\PassOptionsToPackage{sort,compress}{natbib}
\documentclass[final,5p,times,twocolumn,10pt]{elsarticle}

\usepackage{amsmath,amsfonts}
\usepackage{algorithmicx}
\usepackage{algorithm}
\usepackage{array}
\usepackage[caption=false,font=normalsize,labelfont=sf,textfont=sf]{subfig}
\usepackage{textcomp}
\usepackage{stfloats}
\usepackage{url}
\usepackage{verbatim}
\usepackage{siunitx}
\usepackage{cleveref}
\usepackage{xcolor}
\usepackage[acronym]{glossaries}
\usepackage{graphicx}
\usepackage{amssymb}
\usepackage{array}
\usepackage{booktabs}
\usepackage{algpseudocode}
\usepackage{setspace}

\usepackage{tikz}
\usepackage{pgfplots}
\usepackage{graphicx}
\usepackage{standalone} %

\pgfplotsset{compat=1.18}
\usepgfplotslibrary{units}
\usetikzlibrary{positioning,intersections, hobby, patterns, calc, decorations.pathmorphing, decorations.markings, shadows, shapes, plotmarks, shapes.multipart}
\usetikzlibrary{fit}
\usetikzlibrary{arrows.meta}

\usepgfplotslibrary{external}

\tikzset{
  external/system call={pdflatex \tikzexternalcheckshellescape --halt-on-error --interaction=batchmode --jobname "\image" "\texsource"}
}
\tikzexternaldisable %

\pgfplotsset{
            /pgfplots/layers/niceLayers/.define layer set={
                        axis background,
                        axis grid,
                        main,
                        axis ticks,
                        axis lines,
                        axis tick labels,
                        axis descriptions,
                        axis foreground
            }{/pgfplots/layers/standard}
}

\pgfplotsset{
            every axis/.append style={
                        set layers=niceLayers,
                        tick label style={font=\scriptsize},
                        clip marker paths=true,
                        line width=1pt,
                        line cap=round,
                        line join=round,
                        tick style={semithick, color=black},
                        legend style={
                        		   font=\scriptsize,
                                    /tikz/every even column/.append style={column sep=2mm},
                                    cells={anchor=west}, %
                        },
                        xmajorgrids,
                        ymajorgrids,
            }
}

\pgfkeys{/pgf/number format/.cd,1000 sep={}}

\newacronym{pu}{PU}{power unit}
\newacronym{f1}{F1}{Formula 1}
\newacronym{cfd}{CFD}{computational fluid dynamics}
\newacronym{mpc}{MPC}{model predictive control}
\newacronym{drs}{DRS}{Drag Reduction System}
\newacronym{mguk}{MGU-K}{motor-generator unit -- kinetic}
\newacronym{mguh}{MGU-H}{motor-generator unit -- heat}
\newacronym{nn}{NN}{neural network}
\newacronym[plural=OCPs, firstplural=optimal control problems]{ocp}{OCP}{optimal control problem}
\newacronym[plural=NLPs, firstplural=nonlinear programs]{nlp}{NLP}{nonlinear program}
\newacronym{kkt}{KKT}{Karush-Kuhn-Tucker}
\newacronym{mpcc}{MPCC}{mathematical program with complementarity constraints}
\newacronym{licq}{LICQ}{linear independence constraint qualification}
\newacronym{mfcq}{MFCQ}{Mangasarian-Fromovitz constraints qualification}
\newacronym{svo}{SVO}{Social Value Orientation}
\newacronym{ibr}{IBR}{iterated best-response}
\newacronym{mltp}{MLTP}{minimum lap time problem}
\newacronym{qss}{QSS}{quasi-steady-state}
\newacronym{em}{EM}{energy management}
\newacronym[plural=GNEPs, firstplural=Generalized Nash Equilibrium Problems]{gnep}{GNEP}{Generalized Nash Equilibrium Problem}
\newacronym[plural=HEVs, firstplural=hybrid-electric vehicles]{hev}{HEV}{hybrid-electric vehicle}
\newacronym{soc}{SOC}{state-of-charge}
\newacronym{ice}{ICE}{internal combustion engine}
\newacronym{ers}{ERS}{engine recovery system}
\newacronym{rl}{RL}{reinforcement learning}
\newacronym{minlp}{MINLP}{mixed-integer nonlinear program}
\newacronym{mdp}{MDP}{Markov decision process}
\newacronym{sac}{SAC}{soft actor-critic}
\newacronym{des}{DES}{Discrete-event simulation}
\newacronym{dp}{DP}{Dynamic programming}

\newtheorem{problem}{Problem}

\renewcommand{\arraystretch}{1.5}

\newcommand{\DeltaEb}{\Delta E_\mathrm{b,all}}
\newcommand{\DeltaEf}{\Delta E_\mathrm{f,all}}
\newcommand{\TW}{\mathrm{TW}}
\newcommand{\TC}{\mathrm{TC}}
\newcommand{\TA}{\mathrm{TA}}
\newcommand{\PS}{\mathrm{PS}}
\newcommand{\bComp}{b_\mathrm{comp}}
\newcommand{\bPS}{b_\mathrm{PS}}
\newcommand{\bIn}{b_\mathrm{inlap}}
\newcommand{\bOu}{b_\mathrm{outlap}}

\newlength{\myfigskip}
\journal{Transportation Engineering}

\begin{document}
\begin{frontmatter}

\title{Benchmarking Learning-Based Formula 1 Race Strategies}

\author[eth]{Giona Fieni\corref{cor1}}
\ead{gfieni@idsc.mavt.ethz.ch}

\author[eth]{Joschua W\"uthrich}

\author[eth]{Marc-Philippe Neumann}

\author[eth]{Mohammad H. Moradi}

\author[eth]{Christopher H. Onder}

\cortext[cor1]{Corresponding author.}
                  
\affiliation[eth]{organization={Institute for Dynamic Systems and Control, ETH Z\"urich},%
            city={8092 Z\"urich},
            country={Switzerland}}
            
\begin{abstract}
This paper presents two complementary frameworks to optimize Formula 1 race strategies, jointly accounting for energy allocation, tire wear and pit stop timing. First, the race scenario is modeled using lap time maps and a dynamic tire wear model capturing the main trade-offs arising during a race. Then, we solve the problem by means of a mixed-integer nonlinear program that handles the integer nature of the pit stop decisions. The same race scenario is embedded into a reinforcement learning environment, on which an agent is trained. Providing fast inference at runtime, this method is suited to improve human decision-making during real races. The learned policy's suboptimality is assessed with respect to the optimal solution, both in a nominal scenario and with an unforeseen disturbance. In both cases, the agent achieves approximately $\qty{5}{\second}$ of suboptimality on $\qty{1.5}{\hour}$ of race time, mainly attributable to the different energy allocation strategy. This work lays the foundations for learning-based race strategies and provides a benchmark for future developments. 
\end{abstract}


\end{frontmatter}

\section{Introduction}
\gls{f1} is the union between sport, technology and human experience. The performance of the driver and the entire team is of paramount importance. Each year, 11 teams take part to the championship, which counts more than 20 races. Each one consists of a sequence of laps, lasting about $\qty{1.5}{\hour}$. The goal is to finish first or with the highest ranking position to score championship points. 

Since 2014, \gls{f1} has been moving to hybrid-electric propulsion \cite{2025F1_sport,2025F1}. The \gls{pu} features a $\qty{1.6}{\liter}$ V6 turbocharged \gls{ice} and an electric \gls{mguk} connected to a battery. The strategical energy deployment can be optimized to exploit the advantage of the hybrid configuration, increasing the efficiency and performance of the powertrain. Teams have to carefully plan the energy consumption, since refueling is no longer permitted and the battery has a finite capacity. Specifically, race engineers from the pit wall decide how much energy to allocate in the form of fuel and battery targets. On top of that, the onboard fuel mass affects the lap time: A car with an empty fuel tank is lighter and thus faster. 

\begin{figure}
\centering
\begin{tikzpicture}
        \node[anchor=south west, inner sep=0] (img) at (0,0)
            {\includegraphics[width=\columnwidth]{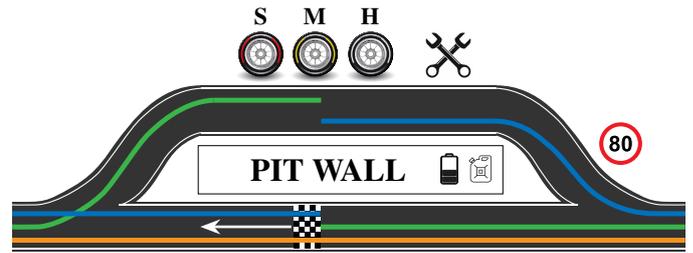}};

        \begin{scope}[x={(img.south east)}, y={(img.north west)}]
            \node at (0.47,0.31) {\large{\textbf{PIT WALL}}};
            \node at (0.37,0.86) {\small{\textbf{S}}};
            \node at (0.45,0.86) {\small{\textbf{M}}};
            \node at (0.532,0.86) {\small{\textbf{H}}};

            \draw[white, -{Stealth[scale=1]},
  		     line width=1pt] 
		     (0.415,0.11) -- (0.28,0.11);
        \end{scope}
    \end{tikzpicture}
\caption{Concept drawing of the start/finish line and pit lane. The engineers in the pit wall decides whether to pit, which compound to mount and how much energy is allocated, in terms of battery and fuel. The available compounds are soft ($S$), medium ($M$) or hard ($H$). In orange, we depict a normal lap, in blue the \textit{inlap} and in green the \textit{outlap}. Furthermore, the pit lane has a velocity limit and the racing direction is given by the arrow.}
\label{fig:Intro}
\end{figure}

Tire degradation is another important factor to account for. Different compounds are available, namely soft ($S$), medium ($M$) and hard ($H$), each one with its own performance and wear characteristics. Once a tire gets too worn, it should be replaced. \Cref{fig:Intro} shows the process of the so-called pit stops. The pit lane runs parallel to the race track with a velocity limit and bypasses the start/finish line. When a pit stop is commanded, at the end of the lap the driver enters the pit lane, slowing down. This is called \textit{inlap} and it is depicted by the blue line. Then the car stops, the crew changes the tires and the driver begins the new lap (green line), called \textit{outlap}. 

The balance between energy allocation, tire wear and pit stops is extremely delicate and complex. For instance, starting the race on hard tires limits the tire wear caused by the higher mass of the full tank, but it initially leads to slower laps. However, it is possible to compensate by allocating more fuel energy, exploiting two joint effects: more energy per lap is available and faster reduction of the car weight. Considering long-term effects, we can intuitively think of a pit stop which lasts $\qty{20}{\second}$, and a car on soft tires gaining $\qty{1}{\second}$ each lap -- the time lost can be recovered in 20 laps. 

We consider decisions taken by the race engineers as ``race strategy''. This includes the combination of energy allocation, pit stop timing and chosen compound. The race strategy is important as much as the car and the driver's performance together. While a good one does not guarantee victory, a suboptimal one can easily lead to a loss of several seconds. A notable example of a successful strategy is the 2004 French Grand Prix. On that occasion, instead of the usual 3-stops strategy, the innovative thinking of Scuderia Ferrari’s engineers led them to adopt an unconventional 4-stops strategy, allowing Michael Schumacher to win despite being hindered by traffic. 

For the race engineers, it is highly important to be able to react to unforeseen events and adapt the strategy as the race deviates from the expectation. Not every scenario can be computed in advance, and the decisions are left to the experience of the engineers. The requirement for rapid and optimal decision-making is of crucial importance. This raises the need of a method that can optimally reject disturbances and at the same time is computationally feasible to be used during a race event.

\subsection{Related work}\label{sec:litRev}
The first part of this review deals with race simulations, optimization and pit stop strategies. Afterwards, we introduce the reader to the literature of tire modeling and wear.

\gls{f1} teams typically rely on Monte Carlo simulations to compute possible race strategies in advance. In the literature, a common approach is to discretize races on a lap-by-lap basis. First attempts of race simulations are present in \cite{bekker2009planning}, where \gls{des} is employed with relatively simple models. Based on that work, \cite{heilmeier2018race} introduces strategic decisions, including the effects of tire degradation, fuel mass, pit stops and overtaking maneuvers. Furthermore, \cite{heilmeier2020application} adds stochastic events using Monte Carlo methods, and in \cite{heilmeier2020virtual,heilmeier2022simulation}, \glspl{nn} are employed to further improve the decision making. 

Regarding race optimization, different problem formulations are needed depending on whether \gls{f1} \cite{duhr2023minimum, bonomi2023evolutionary, heine2023optimization, thomas2025explainable, aguad2024optimizing}, endurance racing \cite{van2022maximum, 10565843, boettinger2023mastering} or Formula E \cite{liu2020formula, liu2021formula} is considered. This is mainly due to the different regulations, vehicles' setups and length of the races. In \cite{duhr2023minimum}, lap time maps are created as the sum of nominal lap times and increases due to battery bounds' proximity. Then, they are embedded within a \gls{nlp} to find the optimal energy allocation between battery and fuel. However, pit stops were not considered as part of the optimization problem. The publication \cite{bonomi2023evolutionary} uses evolutionary algorithms to find race strategies, including pit stop decisions. While the resulting strategies are qualitatively comparable with real racing strategies, heuristics algorithms do not provide formal optimality guarantees. Moreover, battery allocation is not considered, and the discretization of the genetic representation, together with the tuning of mutation parameters, constrains the solution space and impacts scalability. In \cite{heine2023optimization}, \gls{dp} is used to optimize the pit stop strategy in deterministic and stochastic scenarios. While the tire wear and events like safety car are considered, energy management is not taken into account. \gls{dp} and game theory are used in \cite{aguad2024optimizing} to maximize the probability of winning. Neglecting the energy allocation and relying on simple tire degradation models, they set the focus on the game theoretic aspect. The \gls{rl} framework in \cite{thomas2025explainable} does not model energy allocation and the reward is not pure race time, which may dilute the original optimization objective. In \cite{boettinger2023mastering}, the application of \gls{rl} in Gran Turismo racing also relies on reward shaping and a restricted action space, while fuel consumption is only used to extrapolate lap time variations in a non-hybrid powertrain. Due to the different regulations of endurance racing, \cite{van2022maximum, 10565843} optimize pit stops for the battery charging time, rather than due to the tire wear. In Formula E, where pit stops are absent, race strategy is addressed using \gls{rl} in \cite{liu2020formula, liu2021formula}, with a focus on battery management and driving behavior. 

The literature on tires covers modeling, degradation and optimal tire usage. The famous Pacejka's ``magic formula'' \cite{bakker1989new} proposes a detailed model in terms of tire forces, but it does not consider tire wear. The models in \cite{farroni2017physical, sakhnevych2024tyre} describe the thermal behavior and the tire-road interaction by means of differential equations. However, given the level of detail, they are not suited for lap-by-lap discretization. Similar work can be found in \cite{west2020optimal} specifically for \gls{f1}, where the thermal management of tires is optimized. A wear model based on the thermal behavior is studied in \cite{tremlett2016optimal} also for \gls{f1} cars, while \cite{napolitano2023tire} combines physical and statistical analysis to improve the degradation models. 

\subsubsection{Tire data}\label{subsubsec:tiredata}
In this work, we adopt tire wear models that capture degradation dynamics over the race timescale. In particular, our models are inspired by the approach introduced in \cite{heine2023optimization}, where the tire wear is a dynamic state evolving on a lap-by-lap discretization. The influence of the vertical load is explicitly included, allowing the model to account for variations in vehicle mass throughout the race. Model identification is performed using data from the publicly available GitHub repository released by \cite{bonomi2023evolutionary}, developed in collaboration with Pirelli, the official \gls{f1} tire supplier. We were able to separate the relations
\begin{itemize}
\item tire age and vehicle mass to tire wear in \Cref{subsec:tirewear} and
\item tire wear to additional lap time in \Cref{subsub:additionalTime}.
\end{itemize}
This separation is achieved by removing the effect of the fuel mass on the additional lap time. Its contribution is incorporated in the lap time maps of \Cref{subsec:laptime}. Since not enough data were available for the medium compound, its wear characteristics were heuristically derived from the soft and hard compounds. Nevertheless, tire wear models act as placeholders within the proposed framework. Given the larger dataset available, we use the Bahrain circuit as case study.

\subsection{Research statement}
The literature reveals a clear gap in methods that jointly optimize the energy management and the pit stop strategy, while accounting for tire wear. To the best of the authors' knowledge, a solution to the joint problem remains an open research question. Even in works where pit stop strategies are optimized, the proposed methods are computationally demanding, whereby real time feasibility is achieved only through data-driven or heuristic methods. 

A typical challenge with \gls{rl} approaches is to assess the suboptimality of the learned policy. Usually, models used in \gls{rl} environments are not suited for optimization and vice versa. As a consequence, the literature lacks direct and quantitative assessments of policy suboptimality.

\subsection{Contributions}
To address the research questions, we contribute as follows:

We formulate and solve the minimum race time problem as a \gls{minlp}, where we consider pit stops, compound choice and energy allocation. Moreover, we develop a tire wear model that captures the main trade-offs between the compounds relevant throughout a race. This approach provides accurate optimal solutions.

To enable the practical deployment of optimal strategies during actual races, we train an \gls{rl} agent to solve the same problem. The computational burden is shifted to the training phase, enabling fast inference at runtime. This allows to reject disturbances and effectively supports the human decision-making.

Finally, we benchmark the learning-based policy with the \gls{minlp}. A direct comparison is possible because the model, the environment and the problem formulations are equivalent. The suboptimality in terms of race time shows that the \gls{rl} approach is robust and reliable.

Together, these contributions lay the foundations for future learning-based race strategies. In particular, they provide a pathway to overcome the practical limitations of \gls{mpc} applications with mixed-integer decision variables and multi-agent scenarios.

\subsection{Outline}
This paper is structured as follows: In \Cref{sec:racescenario}, we present the race model, describing the factors that influence the race time the most. In \Cref{sec:minlp}, we formulate the optimization problem as a \gls{minlp} to generate optimal solutions, and the \gls{rl} setup is presented in \Cref{sec:sarl}. We then benchmark the \gls{rl} agent by means of case studies in \Cref{sec:results}. Finally, we conclude the paper in \Cref{sec:conclusion} with an outlook on potential extensions of the presented work.

\section{Race model}\label{sec:racescenario}

In this section, we introduce the race model that we consider for our analysis. By means of a cause-and-effect diagram, we present the system's states, models, and boundaries. We aim to capture the physical and regulational effects that most influence the race strategy, in order to build a model that is suitable for classical optimization and \gls{rl}. We focus on the Bahrain circuit, for which we derive the track-dependent coefficients and corresponding lap time maps.

\begin{figure*}
	\centering
		\includegraphics[width=1.8\columnwidth]{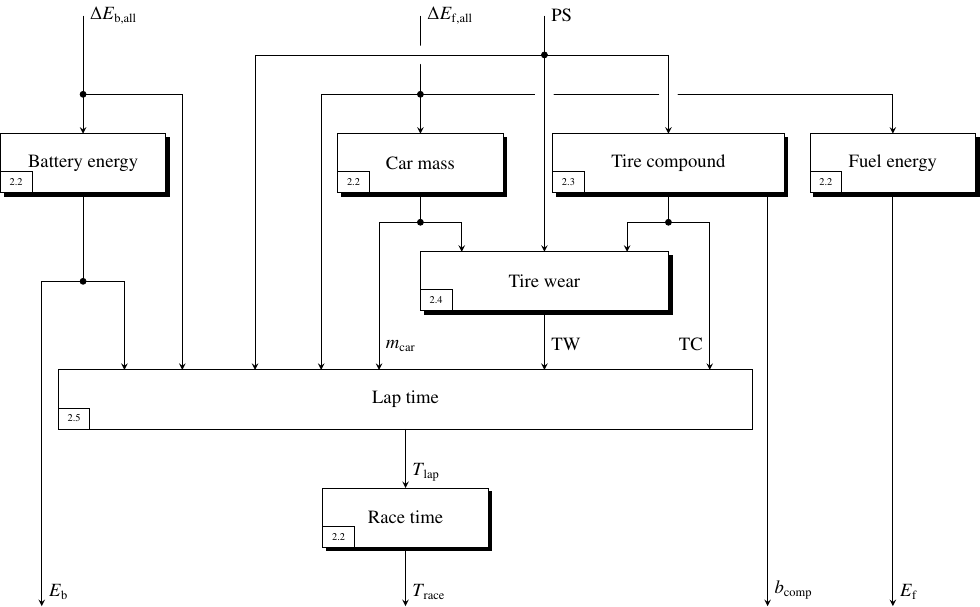}
	\caption{Cause-and-effect diagram for the race scenario. The shaded blocks represent dynamics, while normal blocks are pure algebraic relationships. In the bottom-left corner of each block, we indicate the reference to the section where the system is presented.}
	\label{fig:CausalityDiagram}
\end{figure*}

\subsection{General setup}\label{subsec:setup}
In our problem, the \textit{race time} is the performance metric. For a scenario with only one car, the goal of finishing with the highest position to maximize the championship points cannot be formalized, given the missing information about the competitors. Therefore, the race time is the best measure to maximize achievable points.

The race is discretized on a \textit{lap-by-lap} basis, similar to \cite{duhr2023minimum}. We use the index $k\in\{0, \dots, N_\mathrm{laps}\}$ to address the lap number. A race is a sequence of laps, and their times can be accurately computed in advance by knowing the inputs and states of the system. For instance, it is possible to quantify the additional lap time when allocating less fuel energy. Decisions within each lap--such as racing line and energy management--are beyond the scope of this work and are assumed to be executed in a lap-time-optimal manner, consistent with \cite{ebbesen2017time,salazar2017time,fieni2025game}. The same assumption is employed by \cite{duhr2023minimum}. Indeed, given the available fuel and battery energy as boundary conditions, the lap time optimization problem can be solved to yield the optimal trajectories of MGU-K, engine power and racing line. Mapping each combination of fuel and battery energy targets to a lap time decouples lap and race complexity layers. 

\Cref{fig:CausalityDiagram} showcases the causality of the race model. We start by describing the inputs of the system, i.e., the control variables. They are set at the beginning of lap $k$ and determine the state evolution by the end of the lap. They define the race strategy.
\begin{itemize}
\item The \textit{allocated} battery energy per lap is
\begin{equation}\label{eq:deltaeb}
	\DeltaEb \in [\Delta E_\mathrm{b,min}, \Delta E_\mathrm{b,max}],
\end{equation}
where $\Delta E_\mathrm{b,min} < 0$ and $\Delta E_\mathrm{b,max} > 0$, to deplete or recharge the battery by the end of the lap. 
\item The \textit{allocated} fuel energy per lap is
\begin{align}\label{eq:deltaef}
	\DeltaEf \in & [\Delta E_\mathrm{f,min}, \Delta E_\mathrm{f,max}]\nonumber \\
     = & [0.9, 1.1]\cdot \Delta E_\mathrm{f,nom},
\end{align}
where $\Delta E_\mathrm{f,nom}$ is a nominal fuel load resulting from a strategy with constant fuel allocation. 
\item The \textit{pit stop} decision variable is
\begin{equation}\label{eq:PS}
	\PS\in\{0, 1, 2, 3\},
\end{equation}
where $0$ corresponds to the action ``do not pit'', $1$ to ``pit for soft'', $2$ to ``pit for medium'' and $3$ to ``pit for hard''. Soft, medium, and hard are the tire compounds available for a race, presented in \Cref{subsec:tirecompound}.
\end{itemize}
 
The outputs of the system are
\begin{itemize}
\item the race time $T_\mathrm{race}$, used as performance metric,
\item the battery and fuel energy $E_\mathrm{b}$ and $E_\mathrm{f}$, to comply with energy targets, and
\item the compound change variable $\bComp$, to comply with the regulation of at least one compound change during the race.
\end{itemize}
Intermediate variables are the car's mass $m_\mathrm{car}$, the tire wear $\TW$, the tire compound $\TC$ and the lap time $T_\mathrm{lap}$. 

\subsection{Physical states}\label{subsec:states}
The physical states of the system are the battery energy $E_\mathrm{b}$, the available fuel energy $E_\mathrm{f}$, the car's mass $m_\mathrm{car}$ and the race time $T_\mathrm{race}$. Their update equations are defined as
\begin{align}
E_\mathrm{b}[k+1] &= E_\mathrm{b}[k] + \DeltaEb [k],\label{eq:states1}\\
E_\mathrm{f}[k+1] &= E_\mathrm{f}[k] - \DeltaEf [k],\label{eq:states2}\\
m_\mathrm{car}[k+1] &= m_\mathrm{car}[k] - \frac{\DeltaEf [k]}{H_\mathrm{lhv}},\label{eq:states3}\\
T_\mathrm{race}[k+1] &= T_\mathrm{race}[k] + T_\mathrm{lap} [k],\label{eq:states4}
\end{align}
with $H_\mathrm{lhv}$ being the lower heating value of the fuel.
The battery has a finite capacity and it is fully charged at the race start, resulting in 
\begin{align}
& E_\mathrm{b}[0] = E_\mathrm{b,max},\label{eq:batBC}\\
0 \le & E_\mathrm{b}[k] \le E_\mathrm{b,max}.\label{eq:batBound}
\end{align}
The available fuel energy depends on the mass of fuel loaded on the car prior to the race $m_\mathrm{f,race}$ as
\begin{align}
E_\mathrm{f}[0] = m_\mathrm{f,race}\cdot H_\mathrm{lhv},\label{eq:fuelBC},\\
\end{align}
and the remaining fuel energy cannot be negative:
\begin{align}
E_\mathrm{f}[k] \ge 0.\label{eq:fuelBound}
\end{align}
The initial car mass is 
\begin{equation}\label{eq:massBC}
m_\mathrm{car}[0] = m_\mathrm{car,empty} + m_\mathrm{f,race},
\end{equation}
where $m_\mathrm{car,empty}$ is the weight of the car with an empty tank. With this formulation, it implicitly follows that this state's lower bound is $m_\mathrm{car,empty}$, because we can only consume fuel mass. 
Eventually, the race time is initialized as
\begin{equation}\label{eq:TraceBC}
T_\mathrm{race}[0] = 0.
\end{equation}

\subsection{Tire compound}\label{subsec:tirecompound}
During a race, three types of tire compounds are available: soft (S), medium (M) and hard (H). Each one provides a trade-off between grip and durability. The soft compound provides the most grip, allowing for higher cornering velocity and faster laps, at the cost of higher degradation. Hard tires are the opposite, providing less grip but also less proneness to wear. The medium compound is in between in terms of performance. Once the tires are too worn, they should be replaced during a pit stop. 
The resulting model is 
\begin{equation}\label{eq:TC}
    \TC[k+1] =
    \begin{cases}
        \PS[k], & \text{if} \ \PS[k] > 0,\\
        \TC[k], & \text{otherwise,}
\end{cases}
,
\end{equation}
meaning that the tire compound $\TC$ remains the same if no pit stop is performed, while being updated accordingly to the chosen compound otherwise. At the beginning of the race, we have 
\begin{equation}\label{eq:TCBC}
\TC[0] = \TC_\mathrm{init},
\end{equation}
where $\TC_\mathrm{init}$ is the initial tire compound. 

To define the race strategy, we need the mapping  
\begin{equation}\label{eq:mapping}
\mathcal{L} : \{ 1, 2, 3 \}  \rightarrow \{ S, M, H \}
\end{equation}
which associates numerical code with the corresponding tire compound. 
Let 
\begin{equation}\label{eq:indexPS}
\mathcal{I} = \{ i\in\{0,\dots,N_\mathrm{laps}\} \ | \ \PS[i]>0 \} 
\end{equation}
be the \textit{ordered} set of lap indices at which a pit stop occurred. Then, the race strategy can be defined as the sequence
\begin{equation}\label{eq:strategy}
\mathcal{S} = \left( \mathcal{L}(\TC[i])_i \right) \quad \text{with} \quad i\in \mathcal{I}.
\end{equation}
For example, $\mathcal{S} =(S_0,M_{20},S_{40})$ denotes a soft-medium-soft strategy, where tires were changed at laps 20 and 40.

The regulations impose at least one compound change before the end of the race. To this end, we define the variable  
\begin{equation}\label{eq:b_compound}
    \bComp[k+1] =
    \begin{cases}
        \bComp[k] + 1, & \begin{aligned}
            	&\text{if } \PS[k] > 0 \\
		&\text{and } \PS[k] \neq \TC[k],
        	       \end{aligned}  \\
        \bComp[k], & \text{otherwise,}
    \end{cases}
\end{equation}
which tracks if at any point in the race at least two different compounds were employed. For instance, if at lap $k^{\star}$, the tire compound is $S$, it means that $\TC[k^{\star}] = 1$. If at this lap we decide to pit for $H$, this results in $\PS[k^{\star}] = 3$. Then, $\bComp[k^{\star} + 1] = 1$. The initial and final conditions are 
\begin{align}
\bComp[0] &= 0, \label{eq:bcompBC1}\\
\bComp[N_\mathrm{laps}]& \ge 1,\label{eq:bcompBC2}
\end{align}
We point out that this regulation does not forbid using a compound again. For instance, a $(M_{0},M_{30})$ strategy is not permitted, while a $(M_0,M_{23},S_{38})$ is admissible.

\subsection{Tire wear}\label{subsec:tirewear}
Tire wear is a complex phenomenon. Closely related to the abrasion of the rubber, there exist several types and causes \cite{sakhnevych2024tyre}. The downforce effect typical of \gls{f1} enhances tire abrasion compared to road cars, making rubber wear visible after a few laps. On the one hand, the generated downforce induces high tire forces and stress in the rubber. On the other hand, the downforce effect is less prominent at low velocity, reducing friction force. This increases tire slip, which consumes the tires. Driving style, vehicle setup, wake effects, and temperature also influence degradation, but these aspects are neglected in this work.

Tire degradation is a dominant factor during a race. First, it directly correlates with lap time, which we quantify in \Cref{subsec:laptime}. Tires that provide more grip allow for higher acceleration when exiting a corner or higher velocity at the apexes. Indeed, these zones are called \textit{grip-limited} regions and mostly affect the lap time. Additionally, the driver perceives the tire's feedback and chooses trajectories accordingly. Second, when the tires are worn, the corresponding lap time is no longer competitive and they need to be changed. Hence, the pit stop decision must consider the trade-off between the time spent in the pit lane and the lap time gain provided by new tires.

We first introduce the tire age $\TA$, which is the number of laps that a tire has been used. It is defined as 
\begin{align}\label{eq:tw}
&\TA[0] = 0\\
&\TA[k+1] =
 \begin{cases}
	0, & \text{if} \ \PS[k] > 0,\\
	 \TA[k] + 1, & \text{otherwise.}
\end{cases}
\end{align}
Although this variable is not used in any of the models, the terminology expresses concepts more intuitively. For instance, we can compare the tire wear of different tires using the tire age. 

For the tire wear, we consider the following inputs to be relevant in terms of race strategy. 
\begin{itemize}
    \item \textbf{Pit stop:} It resets the tire wear to a fresh tire.
    \item \textbf{Tire compound:} Each compound wears at different rates, with the soft being the one that lasts the least and the hard being the most durable.
    \item \textbf{Track characteristics:} Length, number of corners, mean cornering velocity, braking zones, and many other factors make each circuit unique. We directly incorporate the track's characteristics into the fitting coefficients of \cref{eq:functiontirewear}.
    \item \textbf{Car's mass:} Rolling friction and normal force increase the tire wear \cite{ivanov2016tire,schutte2021tire,fleischer1973energetische}, both of which are directly proportional to the car's mass. Furthermore, during braking and acceleration phases, the sprung mass of the vehicle deforms the rubber, affecting the degradation.
\end{itemize}

The variable $\TW$ captures different types of wear, without distinguishing between them. Our model reads
\begin{align}
&\TW[0] = 0\label{eq:twBC}\\
& \TW[k+1] = \begin{cases}
	0, & \text{if} \ \PS[k] > 0,\\
	f_{j}(\TW, m_{\mathrm{car}}), & \text{otherwise,}
\end{cases}\label{eq:tw}
\end{align}
where $f_{j}$ is the function related to compound $j$ and is defined as
\begin{equation}\label{eq:functiontirewear}
	f_{j} (\TW, m_{\mathrm{car}})= a_{j}\cdot\TW[k] + b_{j}\cdot\frac{m_{\mathrm{car}}[k]}{m_{\mathrm{car}}[0]}+c_{j} ,
\end{equation}
where $a_{j}$, $b_{j}$ and $c_{j}$ are track-dependent coefficients. Based on the tire wear model of \cite{fleischer1973energetische}, the study \cite{schutte2021tire} linearly relates contact forces to the volume of rubber lost due to wear. While the effects of downforce and circuit topology are incorporated into the track-dependent coefficients, our model separates the wear purely caused by the change in mass. Indeed, it can be observed that during races, the same tires mounted on a lighter car last longer. However, due to the lack of data in the literature, we heuristically choose $b_{j}$. A similar approach is used in \cite{heine2023optimization}, where the tire wear coefficient is a function of the fuel mass onboard.

\begin{figure}
	\centering
		\includegraphics[width=\columnwidth]{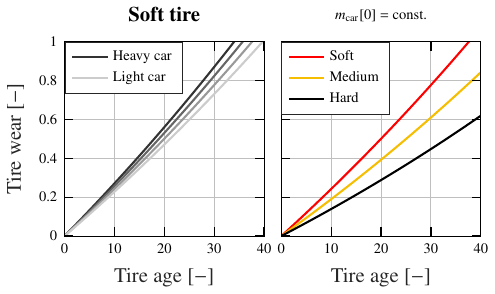}
	\caption{Tire wear as a function of tire age, for different mass of the car on soft tire (left) and a comparison between the compounds given the same mass (right). For confidentiality reasons, the real mass of the car is not indicated.}
	\label{fig:tirewear}
\end{figure}

\Cref{fig:tirewear} shows simulation results of the tire wear model as a function of the tire age. The plot on the left indicates the sensitivity of soft tires w.r.t. the car's initial mass. We notice that with increasing initial $m_\mathrm{car}$ the corresponding tire wear increases marginally. The right plot compares the three compounds, soft, medium, and hard, for the same initial mass. 

\subsection{Lap time}\label{subsec:laptime}
Here, we discuss the factors which determine the lap time. The goal is to provide a model for the lap time based on the conditions and decisions at the beginning of the lap. We assume optimal energy management within the lap and a deterministic pit stop time. The assignment of a probability density function to the duration of pit stops is straightforward but beyond the scope of this paper. Referring to the cause-and-effect diagram of \Cref{fig:CausalityDiagram}, we briefly describe the effect of each input. 
\begin{itemize} 
\item \textbf{Battery energy:} Depending on the initial value, the upper or lower bound might be hit, constraining the battery operation and resulting in slower laps. 
\item \textbf{Allocated battery energy:} The more energy depletion is available within a lap, the faster the lap, and vice versa. 
\item \textbf{Allocated fuel energy:} Same as for the allocated battery energy. 
\item \textbf{Car's mass:} The heavier the car, the less acceleration for the same power output, the slower the lap. 
\item \textbf{Pit stops} influence the current and the next lap. During the inlap, the driver slows down to meet the pit lane velocity limit, allowing to recuperate energy with the \gls{mguk}. The first part of the outlap is still constrained by the velocity limit of the pit lane. 
\item \textbf{Tire wear:} Worn tires provide less grip, resulting in slower cornering velocity and increased lap time. 
\item \textbf{Tire compound:} Each has a different performance for the same tire wear.
\end{itemize}

To model the lap time, we combine a \textit{nominal} lap time with a \textit{correction} term, the latter being a function of tire wear and compound, reading
\begin{align}\label{eq:laptime}
T_\mathrm{lap} = T_\mathrm{nom}(E_\mathrm{b}, \DeltaEb, \DeltaEf, m_\mathrm{car}, \PS) +\Delta T_{j}(\TW),
\end{align}
where $j$ is the compound. We omit the lap index for better readability. While the first term is described by maps, the second is based on existing models fitted on publicly available data as explained in \Cref{subsubsec:tiredata}.   
\subsubsection{Nominal lap time}
The first term is expanded to explicitly state the dependency on $\PS$:
\begin{align}\label{eq:laptimeMap}
T_\mathrm{nom}[k] = 
\begin{cases}
T_\mathrm{lap}[k], &\text{if}\ \PS[k]= 0,\\
T_\mathrm{inlap}[k], &\text{if}\ \PS[k]> 0,\\
T_\mathrm{outlap}[k], &\text{if}\ \PS[k-1]> 0,\\
T_\mathrm{out-inlap}[k], &\text{if}\ \PS[k]> 0 \text{ and } \PS[k-1]> 0.
\end{cases}
\end{align}
This formulation indicates that the inlap map is used when a pit stop occurs at the current lap, the outlap map when a pit stop occurred at the previous lap, and the out-inlap map when two consecutive pit stops are performed. The latter corresponds to the scenario in which the vehicle exits the pit lane and re-enters it within the same lap. 

These maps are produced similarly to \cite{duhr2023minimum}, whose datapoints are generated with a nonlinear lap solver and fitted via neural networks with twice differentiable activation functions. While \cite{duhr2023minimum} relies on a 3D nominal map, outsourcing the battery dependencies $E_\mathrm{b}$ and $\DeltaEb$ to an additive 2D map, we included it directly obtaining a 4D nominal map. In addition to the dedicated inlap and outlap maps, we also added the out-inlap map.

The velocity of the pit lane is restricted by regulations to be \qty{80}{\kilo\meter/\hour}. Hence, in this portion of the lap, we constrain the car's velocity in the inlap and outlap \glspl{nlp}. For the sake of space, we do not show the maps, but we compare the lap times in \Cref{table:laptimes}, given the same conditions of battery, fuel, and mass. 
\begin{table}
\begin{center}
\begin{tabular}{c  c  c  c}
\toprule
\textbf{Nominal lap} & \textbf{Inlap} & \textbf{Outlap} & \textbf{Out-inlap}\\
\midrule
 $\qty{93.1}{\second}$ & $\qty{104.6}{\second}$ & $\qty{108.2}{\second}$ & $\qty{119.7}{\second}$\\
\bottomrule
\end{tabular}
\caption{Comparison of lap times for a nominal lap, an inlap and an outlap, given the same battery, fuel and mass conditions.}\label{table:laptimes}
\end{center}
\end{table}

The differences are explained by the boundary conditions. During inlap, the driver has to slow down only at the end, with the first part of the lap being similar to a nominal one. Additionally, thanks to the deceleration, there is a massive recuperation potential, allowing one to deploy more energy during the lap while meeting the same energy target. On the other hand, during the outlap, the driver starts with the pit lane velocity, slowing down the entire lap.

\subsubsection{Correction term} \label{subsub:additionalTime}
\begin{figure}
	\centering
		\includegraphics[width=\columnwidth]{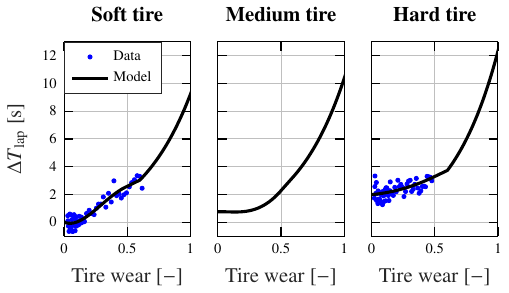}
	\caption{Additional lap time as a function of the tire wear for the three compounds. As mentioned in \Cref{subsubsec:tiredata}, data for medium tires are not available, and the curve is heuristically derived from the soft and hard trends.}
	\label{fig:deltalaptimefit}
\end{figure}

The additional lap time given by the compound and the tire wear is modeled as 
\begin{align}\label{eq:deltalaptime}
\Delta T_{j}[k] = \mathcal{N}_{j}(\TW[k])
,
\end{align}
where $\mathcal{N}$ denotes a twice differentiable fitting function, one for each compound $j$. With this model, we relate the tire wear to a lap time correction. The fitting is shown in \Cref{fig:deltalaptimefit} for soft, medium, and hard tires, where data is available only for $\TW \le 0.6$. The lack of data above this value suggests that tires are usually changed before they are completely worn, since the associated lap time increase makes the car not competitive. We artificially extend the relationship up to $\TW = 1$, to expand the feasible set and avoid a potentially suboptimal constraint. For instance, sacrificing the lap time of a couple of laps towards the end of the race may avoid an additional pit stop. 

\begin{figure}
	\centering
		\includegraphics[width=\columnwidth]{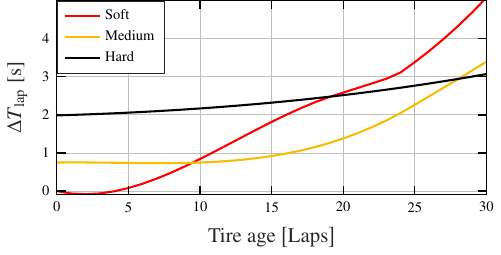}
	\caption{Simulated additional lap time as a function of the tire age, given the same car's mass evolution. These curves result from the concatenation of the models in \Cref{fig:tirewear} and \Cref{fig:deltalaptimefit}.}
	\label{fig:deltalaptime}
\end{figure}

To conclude the modeling part, in \Cref{fig:deltalaptime}, we concatenate and simulate \cref{eq:tw,eq:deltalaptime}, using the same initial mass and fuel consumption. This allows to compare the three different compounds with the same tire age. We note that using fresh soft tires gives $\qty{0}{\second}$ of additional lap time, and using a fresh hard set comes with additional $\qty{2}{\second}$ per lap. On the other side, the expected trade-offs emerge: After 18 laps, the hard compound starts paying off, while the soft tire deteriorates quickly.  

Further model extensions can explicitly include tire temperature as state variable, enabling the capture of two distinct relevant effects: the compound-dependent warm-up time and the additional wear caused by overheating. 
The former can be modeled by incorporating a temperature dependency into \cref{eq:deltalaptime} and resetting the temperature state at each pit stop. The latter can be included in \cref{eq:functiontirewear} as an additional influence factor.

Nevertheless, \cref{eq:deltalaptime} implicitly captures the warm-up effect. Indeed, in \Cref{fig:deltalaptime}, the soft compound shows an inflection during the first 5 laps, reflecting how warmed-up tires perform better than freshly mounted ones. Because the medium tires were heuristically derived from the soft and hard compounds, a similar but less pronounced inflection can be noticed between a tire age of 5 and 10 laps.

\section{Mixed-integer nonlinear program}\label{sec:minlp}

In this section, we state the optimal control problem for the race scenario described in \Cref{sec:racescenario}. The aim is to get an optimization-oriented formulation that is equivalent to the model used for the \gls{rl} environment. In this way, we can directly benchmark the performance of the \gls{rl} agent. Given the presence of logical conditions, its formulation naturally leads to a \gls{minlp}. First, we define auxiliary variables. Then, we reformulate the logical equations. Finally, we state the resulting optimal control problem. 

\subsection{Auxiliary variables}\label{subsec:auxvars}
The only integer variable is $\PS$. However, not every logical condition is only a function of $\PS$, and we need to define auxiliary variables. The first is 
\begin{align}
\bPS[k] = 
\begin{cases}
1, & \text{if } \PS[k] > 0,\\
0, & \text{otherwise,}
\end{cases}
\end{align}
and its equivalent reformulation is  
\begin{equation}\label{eq:bPS}
\bPS = 1 - \frac{1}{6}\cdot (1 - \PS)\cdot(2 - \PS)\cdot(3 - \PS),
\end{equation}
where we dropped the lap index for better readability. 
Furthermore, we also need to determine whether we are in an in- or outlap, described by the variables
\begin{align}
\bIn[k] &= \bPS[k],\label{eq:bLap1}\\
\bOu[k] &= \bPS[k-1],\label{eq:bLap2}\\
\bOu[0] &= 0.\label{eq:bLap3}
\end{align}
To activate the corresponding tire compound-related models, we use the terms
\begin{align}
z_{1}[k] &= 1-\TC[k],\label{eq:Zaux1}\\
z_{2}[k] &= 2-\TC[k],\label{eq:Zaux2}\\
z_{3}[k] &= 3-\TC[k],\label{eq:Zaux3}
\end{align}
whose values, determined by the variable $\TC$, are summarized in \Cref{table:auxvars}.
\begin{table}
\begin{center}
\renewcommand{\arraystretch}{0.99}
\begin{tabular}{c | c c c}
\toprule
 & $z_{1}$ & $z_{2}$ & $z_{3}$\\
\midrule
 $\TC=1$ & $0$ & $1$ & $2$\\
 $\TC=2$ & $-1$ & $0$ & $1$\\
 $\TC=3$ & $-2$ & $-1$ & $0$\\
 \bottomrule
\end{tabular}
\caption{Values taken from the auxiliary variables according to the tire compound.}
\label{table:auxvars}
\end{center}
\end{table}

\subsection{\gls{minlp} reformulation}\label{subsec:refMINLP}
By means of the auxiliary variables, we reformulate all the logical equations needed to define the minimum race time problem. 

\subsubsection{Tire compound}
\Cref{eq:TC} is expressed as
\begin{equation}\label{eq:tcMINLP}
\TC[k+1] = \TC[k] \cdot \left(1 - \bPS[k]\right)+ \PS[k]. 
\end{equation}
which means that if there is a pit stop at lap $k$, we set $\TC$ to $\PS$, while keeping $\TC$ otherwise. Similarly, \cref{eq:b_compound} becomes 
\begin{equation}\label{eq:bcompMINLP}
\bComp[k+1] = \bComp[k] + \left(\PS[k] - \TC[k]\right)^{2}\cdot\bPS[k],
\end{equation}
where the squared term ensures that $\bComp$ can only increase if a new compound is used. Note that this formulation does not exactly match \cref{eq:b_compound} due to the presence of the squared term. Nevertheless, the resulting outcome remains unchanged, and the optimization is not affected. Indeed, as long as the compound is changed, the final condition of \cref{eq:bcompBC2} is fulfilled.

\subsubsection{Tire wear}
The tire wear update \cref{eq:tw} has two logics: one related to $\PS$, the other to $\TC$. To solve this, the three compound functions $f_{j}$ are superposed and activated through the auxiliary variables through
\begin{align}\label{eq:twMINLP}
\TW[k+1] = (1 - \bPS) \cdot &\left(\frac{1}{2}\cdot z_{2}\cdot z_{3} \cdot f_{1}(\TW,m_\mathrm{car})\right.\nonumber\\
& -z_{1}\cdot z_{3} \cdot f_{2}(\TW,m_\mathrm{car})\nonumber\\
&+\left. \frac{1}{2}\cdot z_{1}\cdot z_{2} \cdot f_{3}(\TW,m_\mathrm{car})\right).
\end{align}
If a pit stop takes place, the next tire wear is reset to $0$, otherwise the tire wear equation of the corresponding compound is used.

\subsubsection{Lap time}
The \textit{nominal} lap time maps determine if the current lap is normal, an inlap or an outlap. The selection occurs via the auxiliary variables and it reads 
\begin{align}\label{eq:tnomMINLP}
T_\mathrm{nom} = &(1 - \bIn) \cdot (1 - \bOu)\cdot T_\mathrm{lap} \nonumber\\
& +\bIn \cdot (1 - \bOu)\cdot T_\mathrm{inlap} \nonumber\\
&+ (1 - \bIn) \cdot \bOu\cdot T_\mathrm{outlap} \nonumber\\ 
&+ \bIn \cdot \bOu\cdot T_\mathrm{out-inlap}.
\end{align}
where the lap index is dropped for better readability. For the \textit{correction} term, we use the same superposition of \cref{eq:twMINLP}, resulting in 
\begin{align}\label{eq:tcorrMINLP}
\Delta T = &\frac{1}{2}\cdot z_{2}\cdot z_{3} \cdot \mathcal{N}_{1}(\TW)\nonumber\\
& -z_{1}\cdot z_{3} \cdot \mathcal{N}_{2}(\TW)\nonumber\\
&+\frac{1}{2}\cdot z_{1}\cdot z_{2} \cdot \mathcal{N}_{3}(\TW).
\end{align}

\subsubsection{Car's mass}
We add the explicit constraint 
\begin{equation}\label{eq:carBounds}
m_\mathrm{car,empty}\le m_\mathrm{car}[k]\le m_\mathrm{car,empty} + m_\mathrm{f,race},
\end{equation}
to help the solver's convergence by reducing its search space. This constraint has a purely numerical reason, because it is redundant with the fuel constraints \cref{eq:fuelBC,eq:fuelBound}.

\subsection{Optimal control problem}\label{subsec:MINLP}
We are now ready to formulate the minimum race time \gls{ocp}. 
\begin{problem}
The \gls{ocp} for the race strategy of an \gls{f1} car is
\begin{flalign*}
    \min_{\DeltaEb, \DeltaEf, \PS} T_\mathrm{race}[N_\mathrm{laps}] 
\end{flalign*}
subject to the following constraints:
\begin{alignat*}{2}
		& \text{States:} \quad && \eqref{eq:states1}, \eqref{eq:states2}, \eqref{eq:states3}, \eqref{eq:states4}, \eqref{eq:tcMINLP}, \eqref{eq:bcompMINLP}, \eqref{eq:twMINLP}\\
		& \text{States bounds:} \quad && \eqref{eq:batBound}, \eqref{eq:fuelBound}, \eqref{eq:carBounds} \\
		& \text{Inputs bounds:} \quad && \eqref{eq:deltaeb}, \eqref{eq:deltaef}, \eqref{eq:PS}\\
		& \text{Boundary conditions:} \quad && \eqref{eq:batBC}, \eqref{eq:fuelBC}, \eqref{eq:massBC}, \eqref{eq:TraceBC}, \eqref{eq:TCBC}, \eqref{eq:bcompBC1},\\
		& && \eqref{eq:bcompBC2}, \eqref{eq:twBC}\\
		& \text{Lap time:} \quad && \eqref{eq:laptime}, \eqref{eq:tnomMINLP}, \eqref{eq:tcorrMINLP}\\
		& \text{Auxiliaries:} \quad && \eqref{eq:bPS}, \eqref{eq:bLap1}, \eqref{eq:bLap2}, \eqref{eq:bLap3}, \eqref{eq:Zaux1}, \eqref{eq:Zaux2}, \eqref{eq:Zaux3} \\
	\end{alignat*}  
\end{problem}

We highlight that the problem is already discrete by nature. We parse it with CasADi \cite{andersson2012casadi} and solve it using BONMIN \cite{bonami2012heuristics} with the branch-and-bound algorithm \cite{gupta1985branch}. The computational time ranges from \qty{50}{\second} to \qty{5}{\minute} on a commercial laptop (Apple M2 Max, $\qty{32}{\giga\byte}$ RAM).

\section{Reinforcement learning setup}\label{sec:sarl}
We formulate the race strategy optimization problem as a finite-horizon \gls{mdp}
\begin{equation}
\mathcal{M} = (\mathcal{S}, \, \mathcal{O},\, \mathcal{A},\, T,\, R, \,\mathbb{P}, \,\gamma),
\end{equation}
where $\mathcal{S}$ is the state space, $\mathcal{O}$ is the observation space, $\mathcal{A}$ is the action space, $T$ is the transition function, $R$ is the reward, $\mathbb{P}$ is the transition probability and $\gamma$ is the discount factor. Our \gls{mdp} has discrete decision steps indexed by $k\in\{0, \dots, N_\mathrm{laps}\}$, where each step corresponds to an entire lap. In the following, we present the sets $\mathcal{S}$, $\mathcal{O}$, $\mathcal{A}$, the functions $T$ and $R$ and the transition probability $\mathbb{P}$.
Eventually, we state the Markov property and give details about the implementation.

\subsection{State Space}\label{subsec:ss}
Since \cref{eq:laptimeMap} depends on the input at the previous step $\PS[k-1]$, the update equation \cref{eq:states4} violates the Markov property. To overcome this issue, we define the binary variable 
\begin{equation}
    b_\mathrm{outlap}[k+1] = \begin{cases}
        1, & \text{if } \PS[k]>0,\\
        0, & \text{else,} 
    \end{cases}
\end{equation}
which indicates whether the next lap $k+1$ is an outlap. \Cref{eq:laptimeMap} becomes
\begin{align}\label{eq:laptimeMapRL}
T_\mathrm{nom}[k] = 
\begin{cases}
T_\mathrm{lap}[k], &\text{if}\ \PS[k]= 0,\\
T_\mathrm{inlap}[k], &\text{if}\ \PS[k]> 0,\\
T_\mathrm{outlap}[k], &\text{if}\ b_\mathrm{outlap}[k] = 1,\\
T_\mathrm{out-inlap}[k], &\text{if}\ \PS[k]> 0 \text{ and } b_\mathrm{outlap}[k] = 1,
\end{cases}
\end{align}
recovering the Markov property. At lap $k$, the system is thus described by the state vector
\begin{align}\label{eq:ssRL}
\mathbf{s}_k = & \left(\begin{matrix} E_\mathrm{b}[k] & E_\mathrm{f}[k] & m_\mathrm{car}[k] & T_\mathrm{race}[k] \end{matrix} \right.\nonumber \\
                &  \left. \begin{matrix} b_\mathrm{comp}[k] & \TC[k] & \TW[k] & b_\mathrm{outlap}[k] \end{matrix} \right)
\end{align}
where all the states have already been introduced in \Cref{sec:racescenario}.
We can then define the state space as 
\begin{equation}
    \mathcal{S} = \left\{\mathbf{s}\in \mathbb{R}^8 \mid \mathbf{s} \text{ is feasible in the environment}\right\}.
\end{equation}

\subsection{Observation space}\label{subsec:os}
The environment is fully observable, and we choose the observations to be 
\begin{equation}
    \mathbf{o}_k = \begin{pmatrix}
        \mathbf{s}_k & T_{\mathrm{lap}}[k] & N_\mathrm{laps} - k
    \end{pmatrix},
\end{equation}
where in addition to the states, the agent knows the current lap time and the number of laps remaining. The observation space is 
\begin{equation}
    \mathcal{O} = \left\{\mathbf{o}\in \mathbb{R}^{10} \mid \mathbf{o} \text{ is feasible in the environment} \right\}.
\end{equation}

\subsection{Action Space}
At any step, the agent chooses the action vector
\begin{equation}
    \mathbf{a}_k = \begin{pmatrix} \mathrm{F}[k] & \mathrm{B}[k] & \PS[k] \end{pmatrix}, 
\end{equation}
where 
\begin{itemize}
    \item $\mathrm{F}$ is a normalized fuel energy allocation,
    \item $\mathrm{B}$ is a normalized battery energy allocation,
    \item $\PS$ is the pit stop action as explained in \cref{eq:PS}.
\end{itemize}
The resulting action space is 
\begin{align}
    \mathcal{A} = \left\{ \mathbf{a}\in \mathbb{R}^3 \mid \mathrm{F}\in[0,1], \; \mathrm{B}\in[-1,1], \; \PS\in\{0,1,2,3\} \right\}.
\end{align}

Except for $\PS$, the actions $\mathrm{F}$ and $\mathrm{B}$ must be mapped to the race model inputs $\Delta E_\mathrm{f,all}$ and $\Delta E_\mathrm{b,all}$. To ensure that the policy fully exploits the available action range, we apply element-wise clipping before devising a linear mapping to the physical action space. We define the function 
\begin{align}
    g(a) = g_{a,\mathrm{slope}}\cdot \text{clip} (a, b_\mathrm{min}, b_\mathrm{max}) + g_{a,\mathrm{offset}},
\end{align}
where $a\in\{F,B\}$, $g_{a,\mathrm{slope}}$ and $g_{a,\mathrm{offset}}$ are the linear coefficients and $b_\mathrm{min}$ and $b_\mathrm{max}$ determine the clipping interval. This results into 
\begin{align}
    g(F): [0, 1]& \rightarrow [\Delta E_\mathrm{f,min}, \Delta E_\mathrm{f,max}], \\
    g(B): [-1, 1]& \rightarrow [\Delta E_\mathrm{b,max}, \Delta E_\mathrm{b,min}].
\end{align}
We inverted the mapping of the battery bounds to always have positive actions when energy is being consumed. This improves the convergence during training due to the monotonous relations between energy actions and energy states. The resulting intermediate agent's actions for fuel and battery are thus
\begin{align}
\Delta \tilde{E}_\mathrm{f,all} & = g(F),\\
\Delta \tilde{E}_\mathrm{b,all} & = g(B).
\end{align}

\subsubsection{Backward reachable set}
State constraint satisfaction is obtained by overwriting $\Delta \tilde{E}_\mathrm{f,all}$ and $\Delta \tilde{E}_\mathrm{b,all}$ in case of violation. In addition, it is optimal to finish the race with no energy left, neither in the battery nor in the tank. To this end, the agent's actions are overwritten to ensure that the energy states always lie in a backward reachable set. This set is defined at each lap $k$ by:
\begin{itemize}
\item the maximum remaining energy that can still be fully depleted over the remaining laps
\begin{align}
E_{\mathrm{b,max},k} & = \Delta E_\mathrm{b,max}\cdot (N_\mathrm{laps} - k),\\
E_{\mathrm{f,max},k} & = \Delta E_\mathrm{f,max}\cdot (N_\mathrm{laps} - k),
\end{align}
\item and the minimum amount of fuel energy required to finish the race 
\begin{align}
E_{\mathrm{f,min},k} & = \Delta E_\mathrm{f,min}\cdot (N_\mathrm{laps} - k).
\end{align}
\end{itemize}
For the battery, the race input is overwritten to avoid:
\begin{itemize}
\item Exceeding the upper battery bound $E_\mathrm{b,max}$, 
\item depleting the battery more than \qty{0}{\mega\joule}, and
\item having more battery energy left than the maximum we can allocate for the rest of the race $E_{\mathrm{b,max},k}$,
\end{itemize}
resulting in
\begin{align}
     \Delta E_\mathrm{b,all}[k] = 
     \begin{cases}
        E_\mathrm{b,max} - E_\mathrm{b}[k], & \text{if} \ E_\mathrm{b}[k] + \Delta \tilde{E}_\mathrm{b,all}>E_\mathrm{b,max},\\
        -E_\mathrm{b}[k], & \text{if} \ E_\mathrm{b}[k] + \Delta \tilde{E}_\mathrm{b,all}< \qty{0}{\mega\joule},\\
        E_{\mathrm{b,max},k}  - E_\mathrm{b}[k] , & \text{if} \ E_\mathrm{b}[k] + \Delta \tilde{E}_\mathrm{b,all}>E_{\mathrm{b,max},k}, \\
        \Delta \tilde{E}_\mathrm{b,all} , & \text{otherwise.}
    \end{cases} 
\end{align}
For the fuel, the situations to avoid are
\begin{itemize}
\item when more fuel energy is left than the maximum we can allocate for the rest of the race $E_{\mathrm{f,max},k}$, and
\item when an excessive amount of fuel is consumed, even the minimal fuel allocation for the remaining laps will not result in enough fuel to finish the race,
\end{itemize} 
which are formalized as
\begin{align}
    \Delta E_\mathrm{f,all}[k] =
    \begin{cases}
        E_\mathrm{f}[k] - E_{\mathrm{f,max},k}, & \text{if} \ E_\mathrm{f}[k] - \Delta \tilde{E}_\mathrm{f,all}>E_{\mathrm{f,max},k}, \\
        E_\mathrm{f}[k] - E_{\mathrm{f,min},k}, & \text{if} \ E_\mathrm{f}[k] - \Delta \tilde{E}_\mathrm{f,all}<E_{\mathrm{f,min},k}, \\
        \Delta \tilde{E}_\mathrm{f,all} , & \text{otherwise.}
    \end{cases} 
\end{align}
While the upper bounds defined by $E_{\mathrm{f,max},k}$ and $E_{\mathrm{b,max},k}$ ensure the \textit{optimality} of the problem by consuming all the available energy, the lower bound defined by $E_{\mathrm{f,min},k}$ secures its \textit{feasibility}.

Rather than reducing the feasible action space $\mathcal{A}$, these corrections and overwritings maintain it unchanged and not state-dependent. Additionally, they are implemented as part of the environment dynamics and therefore belong to the transition function $T$, not to the policy. The agent always acts in a fixed, state-independent action space.

\subsection{State Transition Dynamics}
Given the current state $\mathbf{s}_k$ and action $\mathbf{a}_k$, the next state $\mathbf{s}_{k+1}$ is obtained by the deterministic transition function 
\begin{equation}
    T: \mathcal{S}\times\mathcal{A}\rightarrow\mathcal{S},
\end{equation}
such that 
\begin{equation}
\mathbf{s}_{k+1} = T(\mathbf{s}_k, \mathbf{a}_k), 
\end{equation}
where all the state transitions were introduced in \Cref{sec:racescenario} and \Cref{subsec:ss}.

Finally, the episode terminates when no laps remain, i.e.,
\begin{equation}
\text{Done} =
\begin{cases}
1, & \text{if} \ k = N_{\mathrm{laps}},\\
0, & \text{otherwise.}
\end{cases}
\end{equation}

\subsection{Reward Function}
The goal is to minimize the total race time. We therefore define a per-step reward that penalizes long lap times,
\begin{align}
R(\mathbf{s}_k, \mathbf{a}_k, \mathbf{s}_{k+1}) & = r_k \nonumber\\
& = T_{\mathrm{lap,const}} - T_{\mathrm{lap}}(k),
\end{align}
where $T_{\mathrm{lap,const}}$ is a positive constant offset chosen such that the reward remains bounded and numerically well-scaled. Equivalently, maximizing the cumulative return
\begin{equation}
J = \mathbb{E}\!\left[\sum_{k=0}^{N_\mathrm{laps}} \gamma^k\cdot  r_k \right]
\end{equation}
with discount factor $\gamma = 1$ corresponds to minimizing the overall race time.

\subsection{Markov Property}
By construction, the process satisfies the Markov property. Formally, the transition probability satisfies
\begin{equation}
\mathbb{P}\big(\mathbf{s}_{k+1} \mid \mathbf{s}_0, \dots, \mathbf{s}_k,
\mathbf{a}_0, \dots, \mathbf{a}_k\big)
=
\mathbb{P}\big(\mathbf{s}_{k+1} \mid \mathbf{s}_k, \mathbf{a}_k\big),
\end{equation}
for all $k$. This holds because
\begin{itemize}
    \item The state $\mathbf{s}_k$ explicitly contains all quantities that influence future evolution as for \cref{eq:ssRL}.
    \item All deterministic update rules depend only on the current state $\mathbf{s}_k$ and action $\mathbf{a}_k$, including the backward-reachability of fuel and battery.
    \item The reward $r_k$ is a function of $T_{\mathrm{lap}}[k]$ only which, in turn, depends solely on current states and actions.
\end{itemize}
Therefore, given the current states and actions, neither the transition to the next state nor the reward depends on the earlier history, and the environment can be treated as a Markov decision process suitable for standard \gls{rl} algorithms. 

\subsection{Implementation details}
To comply with the regulation of using at least two different compounds, we enforce a compound change if $b_\mathrm{comp}=0$ within the final 20 laps. It is then left to the agent to avoid being corrected and find better strategies. 
The actor and critics neural networks have a multi-headed architecture. One head deals with continuous actions, i.e., fuel and battery allocation, while the other one handles the discrete pit stop action. 

We employ a \gls{sac} algorithm \cite{haarnoja2018soft,haarnoja2018soft2} given its superior performance when dealing with discrete \cite{delalleau2019discrete,christodoulou2019soft} and hybrid action spaces \cite{campos2022soft}. The training of the agent takes approximately $\qty{4}{\hour}$ on a commercial laptop (Apple M2 Max, $\qty{32}{\giga\byte}$ RAM), and the tunable hyperparameters are summarized in \Cref{tab:hyper}.
\begin{table}
\begin{center}
\begin{tabular}{l  c}
\toprule
\textbf{Hyperparameter} & \textbf{Value} \\
\midrule
Learning rate actor & 3e-4\\
Learning rate critic & 1e-3\\
Optimizer & adam \\
Continuous entropy weight & 2\\
Continuous target entropy & 0.1\\
Discrete entropy weight & 3\\
Discrete target entropy & 0.1\\
\bottomrule
\end{tabular}
\caption{Hyperparameters used for the \gls{sac} framework.}\label{tab:hyper}
\end{center}
\end{table}

\section{Benchmarking the RL agent}\label{sec:results}
In this section, we compare the race strategies of the \gls{minlp} and \gls{rl} agent. First, we assess the suboptimality of the \gls{rl} agent against the optimal solution by analyzing the differences in inputs and states. Afterwards, we showcase the ability of the agent to adapt the strategy against an unexpected disturbance. As performance metric we use the race time difference
\begin{equation}\label{eq:racetimediff}
    \Delta T_\mathrm{race} = T_\mathrm{race} - T^\mathrm{MINLP}_\mathrm{race}.
\end{equation}
For brevity, the \gls{minlp} solution is occasionally referred to as the optimal solution. 
	
\subsection{Nominal case}\label{subsec:resultsNom}
\begin{figure}
	\centering
		\includegraphics[width=\columnwidth]{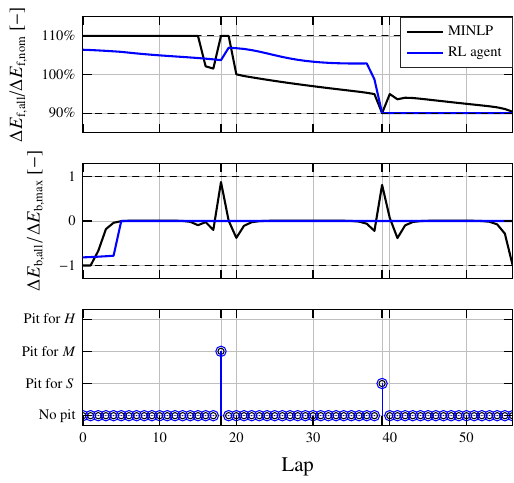}
	\caption{Control variables and actions of the \gls{minlp} and the \gls{rl} agent. From top to bottom: fuel energy allocation, battery energy allocation, pit stop decision.}
	\label{fig:actionsbenchmarkinputs}
\end{figure}

\begin{figure}
	\centering
		\includegraphics[width=\columnwidth]{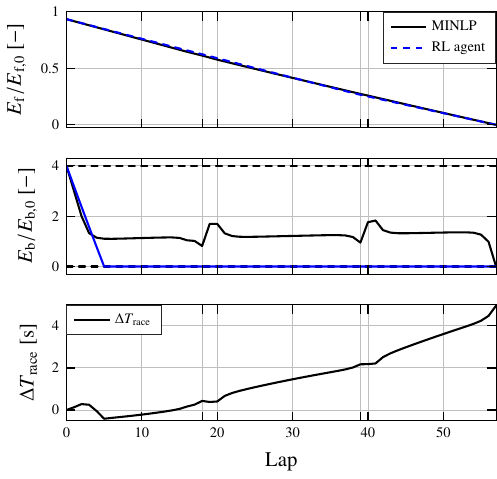}
	\caption{Subset of states resulting from the \gls{minlp} and the \gls{rl} agent. From top to bottom: (normalized) fuel energy, (normalized) battery energy and race time difference.}
	\label{fig:actionsbenchmarkstates}
\end{figure}

Here, we directly compare the \gls{rl} agent's race strategy with the \gls{minlp} solution. Since the model for the optimization problem and the environment are equivalent, we can rely on a precise benchmark. We point out that the initial conditions are the same for both problems, and both strategies start on medium tires. The control variables/actions are shown in \Cref{fig:actionsbenchmarkinputs}, while \Cref{fig:actionsbenchmarkstates} shows a subset of the states and the race time difference. 

The resulting suboptimality is $\qty{4.96}{\second}$, which corresponds to $\qty{0.09}{\%}$ of the total race time. We highlight that all fuel and battery limits are respected, and there is no energy left at the end of the race. The pit stop strategy perfectly reflects the $(M_0,M_{18},S_{39})$ choice of the optimal solution, and the suboptimality is due to the difference in energy allocation. Although the overall trend is similar, the \gls{rl} agent fails to describe the fine adjustments of battery and fuel energy around the pit stops.

Analyzing the overall trend, the \gls{minlp} solution allocates more fuel and battery energy for the first part of the race. Consuming fuel lightens the car which, in turn, makes it faster. Thus, the goal is to fast reduce its weight, to profit for more laps of the reduced mass. Since this process requires some laps, more battery energy is used to initially compensate for the weight of the full tank. The agent is able to capture these trends of fuel and battery strategies by considering the change in mass of the car.

For the battery energy management around the pit stops, we notice the following: The optimal solution shows an increased battery usage before and after the pit stops, and a considerable recharge during the inlap. Here, the driver decelerates the car from $\qty{300}{\kilo\meter/\hour}$ to the pit lane speed limit at $\qty{80}{\kilo\meter/\hour}$. This results in a massive recuperation potential. The harvested energy is used before and after the pit stop to compensate for the time spent in the pit lane. Except for the first 5 laps, the \gls{rl} agent chooses a charge-sustained strategy, even around pit stops. The sensitivity of the race time given by the different battery allocation around pit stops is too small for the RL agent to learn it.

The agent keeps a battery level of $E_\mathrm{b}=0$ during the race, while the \gls{minlp} chooses to stay around $E_\mathrm{b}=1.2$ in normalized units. We point out that this is just the battery level at the start/finish line. Given the characteristics of the Bahrain circuit, starting the lap with an empty battery is not detrimental in terms of lap time. Indeed, the lap time maps are flat in the region close to the lower battery bound (not shown here), and the gradients are too small for the agent to notice this trend. This is a consequence of the exploration-exploitation trade-off commonly seen in \gls{rl}.

\Cref{table:compTime} summarizes the computational burden needed to evaluate the strategy for the entire race. Despite the small differences in energy allocation, the \gls{rl} race strategy $(M_0,M_{18},S_{39})$ is computed in less than a second. This feature is particularly interesting, and we showcase its potential in the case study below.

\begin{table}
\begin{center}
\begin{tabular}{c  c  c}
\toprule
 & \textbf{\gls{minlp}} & \textbf{\gls{rl} agent}\\
\midrule
 Computational time & $\qty{55}{\second}$ & $<\qty{1}{\second}$\\
\bottomrule
\end{tabular}
\caption{Computational time required by the \gls{minlp} solver and the \gls{rl} agent to obtain the solution for the entire race.}\label{table:compTime}
\end{center}
\end{table}

\subsection{Scenario with unexpected disturbance}\label{subsec:resultsDist}
\begin{figure}
	\centering
		\includegraphics[width=\columnwidth]{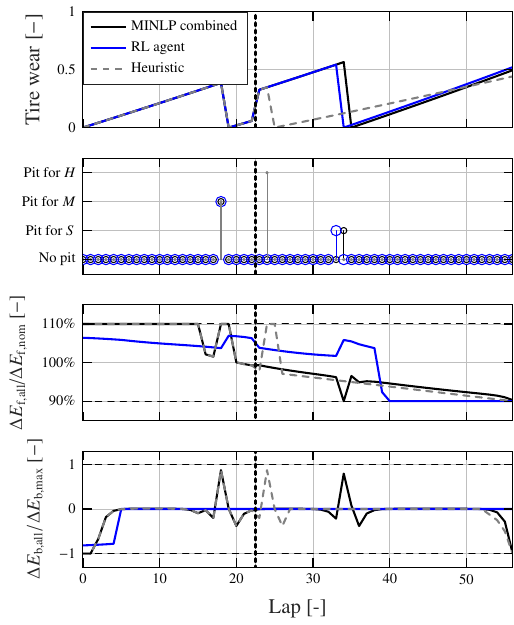}
	\caption{Tire wear, pit stop action and energy allocation (fuel and battery) in case of a disturbance for the three strategies: causal \gls{minlp}, \gls{rl} agent and heuristic. The disturbance happens at lap 22, depicted with a dashed line.}
	\label{fig:disturbance}
\end{figure}

In this scenario, we investigate the case where a sudden increase in tire wear forces the strategy to change. In \gls{f1}, laps last longer than $\qty{1}{\minute}$. Usually, multiple pit stops strategies are defined in advance by race engineers, but during a race there are multiple sources of disturbance. Recomputing the strategies in real time is a difficult task, and the decision-making is left to the experience of race engineers. Being able to optimally adapt these strategies is crucial to win the race. For instance, a typical situation is when the driver runs oﬀ track or must brake hard to avoid a crash. We simulate it by artificially increasing the tire wear during lap $22$. \Cref{fig:disturbance} shows tire wear, pit stops and energy allocation for the three strategies below. Additionally, \Cref{table:disturbance} summarizes the suboptimality in terms of race time and the corresponding computational time.

\begin{itemize}
    \item Combined \gls{minlp} solution $(M_0, M_{18}, S_{34})$. We follow an optimal strategy until lap $22$, and then we optimize again from lap $22$ to the end of the race. This combination results in a \textit{causal} solution with which the other strategies can be benchmarked. 
    \item RL strategy $(M_0, M_{18}, S_{33})$. Since the agent is evaluated in simulation, it naturally adapts its strategy given the sudden increase in tire wear. 
    \item Heuristic strategy $(M_0, M_{18}, H_{24})$. Here, we simulate the conservative decision of a race engineer to pit for hard tires immediately after the disturbance, in order to avoid 3 pit stops. As energy management, we adapt fuel and battery allocation around the pit stops to be similar to the \gls{minlp} version.
\end{itemize}

Before the disturbance (vertical dashed line), the \gls{minlp} and \gls{rl} solution coincide with those in \Cref{subsec:resultsNom}: Both pit for medium tires at lap $18$, with the same energy allocation difference. The heuristic solution exactly reflects the \gls{minlp} solution: We imitate an engineer that precomputed a race strategy and executes it. 

After the disturbance, the \gls{minlp} and the \gls{rl} agent change their pit stop strategy almost identically. With only a difference of 1 lap, both pit for soft tires earlier than in the previous case study, showing that the agent adapts to unforeseeable situations. The major difference remains in the energy allocation. By the end of the race, the agent has $\qty{5.08}{\second}$ of additional time. Recalling the suboptimality of $\qty{4.96}{\second}$ from \Cref{subsec:resultsNom}, we conclude that the main source of suboptimality is caused by the energy allocation. 

The agent shows its superiority by providing close-to-optimal pit stop strategy with a negligible computational time. Since the process is Markovian, the agent naturally adapts to the states of lap $k$ without additional overhead, regardless of the presence of disturbances. This property is of particular interest, because during a race, relying on optimization techniques with unpredictable computational times is impractical. 

To evaluate its performance, we consider a typical \gls{f1} situation. After the disturbance happens, a decision under pressure has to be taken, and the race engineer decides for the so-called ``go long'' strategy. Relative to the \gls{minlp} solution, this strategy is $\qty{31.45}{\second}$ slower, against the $\qty{5.08}{\second}$ of the agent, emphasizing its advantage.

We neglected external factors in this analysis. For instance, the vehicle's stability could be compromised or damages to the vehicle's body may require an immediate pit stop. In these cases, only the judgement of race engineers is relevant. Nevertheless, the \gls{rl} solution can provide valuable predictions and insights for better-informed decisions of race engineers. The agent is suited for online deployment for disturbance rejection, as the evaluation of a feedforward network comes with minimal computational overhead.

\begin{table}
\begin{center}
\begin{tabular}{l  c  c}
\toprule
 & $\Delta T_\mathrm{race}$ & \textbf{Computational time}\\
\midrule
Causal \gls{minlp} & $\qty{0}{\second}$ & $\qty{68}{\second}$\\
\gls{rl} agent & $\qty{5.08}{\second}$ & $<\qty{1}{\second}$\\
Heuristic & $\qty{31.45}{\second}$ & $-$\\
\bottomrule
\end{tabular}
\caption{Race time difference and computational time for the causal \gls{minlp} solution, the \gls{rl} agent and the heuristic in case of a disturbance. The computational time of the heuristic is not reported, as it represents the conceptual thinking of a race engineer.}\label{table:disturbance}
\end{center}
\end{table}

\section{Conclusion and outlook}\label{sec:conclusion}
In this paper, we filled the literature gap of race strategies, where energy allocation, pit stop and tirere jointly considered. We show how the same problem can be solved by means of a \gls{minlp} and \gls{rl}. This way, we obtain almost identical results with completely different goals. The \gls{minlp} serves as a ground-truth benchmark for the \gls{rl} agent, and delivers the optimal solution. In a first case study, the \gls{rl} agent suboptimality is $\qty{0.09}{\percent}$, but with fast inference. This property is particularly useful during a race, where decisions have to be taken within a few seconds and there is not enough time to recompute an optimal solution. To this end, we show in the second case study how the \gls{rl} agent reacts to an unexpected event. In addition to the causal optimal solution, we also compare it to a heuristic simulating the decision of a race engineer. The results show that the \gls{rl} approach is robust and reliable, with a negligible inference time.  

For future research, we have several ideas. First, we can add probabilistic models, such as pit stop timing, traffic or weather predictions. Additional inputs or models used in real racing may also be integrated. For instance, a target pace could be introduced, allowing the driver to push the car to achieve faster lap times at the expense of increased wear, or to account for brake temperature dynamics as a function of pace. Regulatory constraints such as the minimum car's weight at the end of the race can be included. Given the satisfactory performance of the agent, we can now explore scenarios with multiple agents interacting with each other, whether representing competitors or teammates. This could result in unintuitive strategies that are difficult to predict. Eventually, while the \gls{rl} agent is precise in the pit stop decision, it still struggles to achieve an optimal energy allocation. On the contrary, the \gls{minlp} is not real-time feasible. This motivates the integration of the two approaches by solving a continuous \gls{nlp} parametrized by the \gls{rl} agent's discrete pit stop decisions within an \gls{mpc} framework. In this setting, the computational burden associated with integer variables is handled by the agent, while a continuous optimizer manages the energy allocation. By eliminating integer variables from the online optimization, real-time feasibility can be achieved.

\section*{Acknowledgments}
We would like to express our deep gratitude to Ilse New for her helpful and valuable comments during the proofreading phase. We also thank Fabio Widmer for providing important feedback and Manish Prajapat, Joram Eickhoff and Nazim Yasar for the insightful discussions on reinforcement learning.

\bibliographystyle{elsarticle-num} 
\bibliography{bibliography.bib}

\begin{thebibliography}{10}
\expandafter\ifx\csname url\endcsname\relax
  \def\url#1{\texttt{#1}}\fi
\expandafter\ifx\csname urlprefix\endcsname\relax\def\urlprefix{URL }\fi
\expandafter\ifx\csname href\endcsname\relax
  \def\href#1#2{#2} \def\path#1{#1}\fi

\bibitem{2025F1_sport}
FIA, 2025 {F}ormula {O}ne sporting regulations, Tech. rep., Geneva, Switzerland
  (2025).

\bibitem{2025F1}
FIA, 2025 {F}ormula {O}ne technical regulations, Tech. rep., Geneva,
  Switzerland (2025).

\bibitem{bekker2009planning}
J.~Bekker, W.~Lotz, Planning {F}ormula {O}ne race strategies using
  discrete-event simulation, Journal of the Operational Research Society 60~(7)
  (2009) 952--961.

\bibitem{heilmeier2018race}
A.~Heilmeier, M.~Graf, M.~Lienkamp, A race simulation for strategy decisions in
  circuit motorsports, in: 2018 21st International Conference on Intelligent
  Transportation Systems (ITSC), IEEE, 2018, pp. 2986--2993.

\bibitem{heilmeier2020application}
A.~Heilmeier, M.~Graf, J.~Betz, M.~Lienkamp, Application of monte carlo methods
  to consider probabilistic effects in a race simulation for circuit
  motorsport, Applied Sciences 10~(12) (2020) 4229.

\bibitem{heilmeier2020virtual}
A.~Heilmeier, A.~Thomaser, M.~Graf, J.~Betz, Virtual strategy engineer: Using
  artificial neural networks for making race strategy decisions in circuit
  motorsport, Applied Sciences 10~(21) (2020) 7805.

\bibitem{heilmeier2022simulation}
A.~M. Heilmeier, Simulation of circuit races for the objective evaluation of
  race strategy decisions, Ph.D. thesis, Technische Universit{\"a}t M{\"u}nchen
  (2022).

\bibitem{duhr2023minimum}
P.~Duhr, D.~Buccheri, C.~Balerna, A.~Cerofolini, C.~H. Onder, Minimum-race-time
  energy allocation strategies for the hybrid-electric {F}ormula 1 power unit,
  IEEE Transactions on Vehicular Technology 72~(6) (2023) 7035--7050.

\bibitem{bonomi2023evolutionary}
A.~Bonomi, E.~Turri, G.~Iacca, Evolutionary {F1} race strategy, in: Proceedings
  of the Companion Conference on Genetic and Evolutionary Computation, 2023,
  pp. 1925--1932.

\bibitem{heine2023optimization}
O.~F.~C. Heine, C.~Thraves, On the optimization of pit stop strategies via
  dynamic programming, Central European Journal of Operations Research 31~(1)
  (2023) 239--268.

\bibitem{thomas2025explainable}
D.~Thomas, J.~Jiang, A.~Kori, A.~Russo, S.~Winkler, S.~Sale, J.~McMillan,
  F.~Belardinelli, A.~Rago, Explainable reinforcement learning for {Formula
  One} race strategy, in: Proceedings of the 40th ACM/SIGAPP Symposium on
  Applied Computing, 2025, pp. 1090--1097.

\bibitem{aguad2024optimizing}
F.~Aguad, C.~Thraves, Optimizing pit stop strategies in {Formula} 1 with
  dynamic programming and game theory, European Journal of Operational Research
  319~(3) (2024) 908--919.

\bibitem{van2022maximum}
J.~van Kampen, T.~Herrmann, M.~Salazar, Maximum-distance race strategies for a
  fully electric endurance race car, European Journal of Control 68 (2022)
  100679.

\bibitem{10565843}
J.~van Kampen, M.~Moriggi, F.~Braghin, M.~Salazar, Model predictive control
  strategies for electric endurance race cars accounting for competitors'
  interactions, IEEE Control Systems Letters 8 (2024) 1799--1804.
\newblock \href {https://doi.org/10.1109/LCSYS.2024.3417174}
  {\path{doi:10.1109/LCSYS.2024.3417174}}.

\bibitem{boettinger2023mastering}
M.~Boettinger, D.~Klotz, Mastering {N}ordschleife--{A} comprehensive race
  simulation for {AI} strategy decision-making in motorsports, arXiv preprint
  arXiv:2306.16088 (2023).

\bibitem{liu2020formula}
X.~Liu, A.~Fotouhi, Formula-{E} race strategy development using artificial
  neural networks and monte carlo tree search, Neural Computing and
  Applications 32~(18) (2020) 15191--15207.

\bibitem{liu2021formula}
X.~Liu, A.~Fotouhi, D.~J. Auger, Formula-{E} race strategy development using
  distributed policy gradient reinforcement learning, Knowledge-Based Systems
  216 (2021) 106781.

\bibitem{bakker1989new}
E.~Bakker, H.~B. Pacejka, L.~Lidner, A new tire model with an application in
  vehicle dynamics studies, SAE transactions (1989) 101--113.

\bibitem{farroni2017physical}
F.~Farroni, A.~Sakhnevych, F.~Timpone, Physical modelling of tire wear for the
  analysis of the influence of thermal and frictional effects on vehicle
  performance, Proceedings of the Institution of Mechanical Engineers, Part L:
  Journal of Materials: Design and Applications 231~(1-2) (2017) 151--161.

\bibitem{sakhnevych2024tyre}
A.~Sakhnevych, A.~Genovese, Tyre wear model: a fusion of rubber
  viscoelasticity, road roughness, and thermodynamic state, Wear 542 (2024)
  205291.

\bibitem{west2020optimal}
W.~J. West, D.~J. Limebeer, Optimal tyre management of a {Formula One} car,
  IFAC-PapersOnLine 53~(2) (2020) 14456--14461.

\bibitem{tremlett2016optimal}
A.~Tremlett, D.~Limebeer, Optimal tyre usage for a {Formula One} car, Vehicle
  System Dynamics 54~(10) (2016) 1448--1473.

\bibitem{napolitano2023tire}
G.~Napolitano~Dell'Annunziata, G.~Adiletta, F.~Farroni, A.~Sakhnevych,
  F.~Timpone, Tire wear sensitivity analysis and modeling based on a
  statistical multidisciplinary approach for high-performance vehicles,
  Lubricants 11~(7) (2023) 269.

\bibitem{ebbesen2017time}
S.~Ebbesen, M.~Salazar, P.~Elbert, C.~Bussi, C.~H. Onder, Time-optimal control
  strategies for a hybrid electric race car, IEEE Transactions on control
  systems technology 26~(1) (2017) 233--247.

\bibitem{salazar2017time}
M.~Salazar, P.~Elbert, S.~Ebbesen, C.~Bussi, C.~H. Onder, Time-optimal control
  policy for a hybrid electric race car, IEEE Transactions on Control Systems
  Technology 25~(6) (2017) 1921--1934.

\bibitem{fieni2025game}
G.~Fieni, M.-P. Neumann, F.~Furia, A.~Caucino, A.~Cerofolini, V.~Ravaglioli,
  C.~H. Onder, Game theory in formula 1: From physical to strategic
  interactions, European Journal of Control (2026) 101554.

\bibitem{ivanov2016tire}
R.~Ivanov, Tire wear modeling, Transport problems 11~(3) (2016) 111--120.

\bibitem{schutte2021tire}
J.~Sch{\"u}tte, W.~Sextro, Tire wear reduction based on an extended multibody
  rear axle model, Vehicles 3~(2) (2021) 233--256.

\bibitem{fleischer1973energetische}
G.~Fleischer, Energetische {M}ethode der {B}estimmung des {V}erschlei{\ss}es,
  Schmierungstechnik 4~(9) (1973) 269--274.

\bibitem{andersson2012casadi}
J.~Andersson, J.~{\AA}kesson, M.~Diehl, {CasADi}: A symbolic package for
  automatic differentiation and optimal control, in: Recent advances in
  algorithmic differentiation, Springer, 2012, pp. 297--307.

\bibitem{bonami2012heuristics}
P.~Bonami, J.~P. Gon{\c{c}}alves, Heuristics for convex mixed integer nonlinear
  programs, Computational Optimization and Applications 51~(2) (2012) 729--747.

\bibitem{gupta1985branch}
O.~K. Gupta, A.~Ravindran, Branch and bound experiments in convex nonlinear
  integer programming, Management science 31~(12) (1985) 1533--1546.

\bibitem{haarnoja2018soft}
T.~Haarnoja, A.~Zhou, P.~Abbeel, S.~Levine, Soft actor-critic: {O}ff-policy
  maximum entropy deep reinforcement learning with a stochastic actor, in:
  International conference on machine learning, Pmlr, 2018, pp. 1861--1870.

\bibitem{haarnoja2018soft2}
T.~Haarnoja, A.~Zhou, K.~Hartikainen, G.~Tucker, S.~Ha, J.~Tan, V.~Kumar,
  H.~Zhu, A.~Gupta, P.~Abbeel, et~al., Soft actor-critic algorithms and
  applications, arXiv preprint arXiv:1812.05905 (2018).

\bibitem{delalleau2019discrete}
O.~Delalleau, M.~Peter, E.~Alonso, A.~Logut, Discrete and continuous action
  representation for practical rl in video games, arXiv preprint
  arXiv:1912.11077 (2019).

\bibitem{christodoulou2019soft}
P.~Christodoulou, Soft actor-critic for discrete action settings, arXiv
  preprint arXiv:1910.07207 (2019).

\bibitem{campos2022soft}
G.~Campos, N.~H. El-Farra, A.~Palazoglu, Soft actor-critic deep reinforcement
  learning with hybrid mixed-integer actions for demand responsive scheduling
  of energy systems, Industrial \& Engineering Chemistry Research 61~(24)
  (2022) 8443--8461.

\end{thebibliography}

\end{document}